\documentstyle[aps,prl,epsfig]{revtex}

\mathchardef\ogon="012C%
\newcommand{\as}{a\kern-0.22em\lower.40ex\hbox{$_{\ogon}$}}

\begin{document}
\wideabs{
\title{Creation of a dipolar superfluid in optical lattices}
\author{B. Damski$^{1,2}$, L. Santos$^{1}$, E. Tiemann$^{3}$,
M. Lewenstein$^{1}$, S. Kotochigova$^{4}$,  
P. Julienne$^{4}$ and P. Zoller$^{5}$}
\address{
(1) Institut f\"ur Theoretische Physik, Universit\"at Hannover, D-30167
Hannover, Germany\\
(2) Instytut Fizyki,
  Uniwersytet Jagiello\'nski, Reymonta 4, PL-30 059 Krak\'ow, Poland \\
(3) Institut f\"ur Quantenoptik, Universit\"at Hannover, D-30167
Hannover, Germany\\
(4) National Institute of Standards and Technology, Gaithersburg, 
MD 20899, USA.\\
(5) Institut f\"ur Theoretische Physik, Universit\"at Innsbruck,
 A--6020 Innsbruck, Austria}
\maketitle

\begin{abstract}

We show that by loading a Bose-Einstein condensate (BEC)  of two different atomic 
species into an optical lattice, it is possible to achieve a Mott-insulator phase 
with exactly one atom of each species per lattice site. 
A subsequent photo-association leads to the formation  
of one heteronuclear molecule with a large electric dipole moment, 
at each lattice site. 
The melting of such dipolar Mott-insulator  
creates a dipolar superfluid, and eventually a dipolar molecular BEC. 
\end{abstract}
}

The physics of strongly correlated systems in 
ultra-cold trapped quantum gases has attracted a growing interest recently. 
The experimental progress on the trapping and cooling techniques, 
and the control 
of the inter-atomic potentials by means of Feshbach resonances allow  
to analyze situations beyond the validity of the  mean-field approach.
In particular, the possibility to achieve 1D bosonic gases 
opens the pathways toward a gas of impenetrable bosons \cite{Tonks}. 
Also, a  rapidly rotating trapped bosonic gas should 
exhibit effects similar  
to  the fractional quantum-Hall effect 
\cite{Paredes}. 
The field of strongly correlated atomic systems concerns also  
ultra-cold Bose gases with large scattering 
lengths \cite{Cowell,Claussen}, and the Mott-insulator (MI) to superfluid 
(SF) phase  transition \cite{Fisher} in cold bosonic gases in optical lattices 
\cite{Jaksch98,Goral02}. 
The latter has been recently observed in the remarkable 
experiment~\cite{Greiner}.


The physics of ultra-cold molecular gases  is also a very 
 active research area.
Cooling and trapping of molecules can be achieved
by means of buffer-gas techniques \cite{Doyle}, 
by employing the dipolar moments of polar molecules \cite{Meijer}, 
and by means of 
photo-association of already existing ultra-cold atomic gases 
\cite{Heinzen}. Jaksch {\it et al.} \cite{Jaksch02} 
have recently 
proposed that a bosonic atomic gas, which is in a   
MI phase with exactly two atoms per site  in an optical lattice, 
can be used to 
obtain via photo-association a superfluid gas of homonuclear molecules, 
and eventually a molecular BEC.


The influence of dipole-dipole forces on the properties of ultra-cold 
gases has also drawn a considerable 
attention. 
It has been shown that these forces significantly modify 
the ground state and excitations 
of trapped dipolar BEC's \cite{You,Goral,Santos,Goral01,Meystre,Giovanazzi}. 
In addition, since dipole-dipole interactions can be quite strong 
relative to the short-range (contact) interactions,     
dipolar particles are considered to be promising candidates for the 
implementation 
of fast and robust quantum-computing schemes \cite{DDgates,qcomp}. 
Several sources of cold dipolar bosons have been proposed, including 
atoms with 
large magnetic moments \cite{Goral}, and externally induced electric dipoles 
\cite{You,Santos,DDgates}. Perhaps the most promising perspective 
in this sense is 
provided by ultra-cold gases of polar molecules with large 
electric dipole moments, which could eventually dominate the 
physics of the system 
\cite{Santos,Goral01}.


This Letter is devoted to the analysis of the generation of ultra-cold 
polar molecules in an optical lattice. 
We consider the loading of the lattice by a superfluid of 
two different atomic species, and analyze the transition into a MI 
phase with exactly one atom of each species per site. This transition 
is followed by the formation of polar dimers by photo-association, or adiabatic passage. 
Finally,  the quantum melting of the molecular Mott phase into a polar molecular superfluid 
takes place.

We consider a miscible gas of two trapped atomic bosonic species at very low 
temperature $T$. Although our calculations could in principle be employed 
for arbitrary atoms, we have assumed in the calculations below 
the particular 
mixture of $^{41}{\rm K}$ and $^{87}{\rm Rb}$, whose simultaneous 
Bose-Einstein condensation 
has been recently realized \cite{Inguscio}. We additionally consider 
that the binary 
gas is confined in a 2D optical lattice, similarly to the 3D 
one-component case recently 
observed in Ref. \cite{Greiner}. The transverse dimension is considered as 
harmonically confined. The physics of the system is governed by the 
two-species generalization of the Bose-Hubbard (BH) model, 
described by  the Hamiltonian:
\begin{eqnarray} \label{H}
H&=&\sum_{<i,j>}[J_aa_{i}^{\dagger}a_{j} + J_bb_{i}^{\dagger}b_{j}]
+U_{ab} \sum_{i}n_{ai}n_{bi}
\nonumber \\
&+&\frac{1}{2}\sum_{i}[U_{0a}n_{ai}(n_{ai}-1) +U_{0b}n_{bi}(n_{bi}-1)],
\end{eqnarray}
\noindent where $a_i$, $b_{i}$ are 
the annihilation operators of $^{41}{\rm K}$ and $^{87}{\rm Rb}$ atoms 
at the  
lattice site $i$, which occupy a state described by 
the Wannier functions $w_a({\bf r}-{\bf r}_{i})$, 
$w_b({\bf r}-{\bf r}_{i})$ of the lowest energy band, 
localized on this site. This implies that energies 
involved in the system are small compared to the excitation energies 
to the second band. We denote the position of  
the local minimum of the optical 
potential as ${\bf r}_{i}$, and the number operator of $\rm K$ ($\rm Rb$) 
atoms at the site $i$ as $n_{ai}=a_{i}^{\dagger}a_{i}$
($n_{bi}=b_{i}^{\dagger}b_{i}$). 
In Eq. (\ref{H}) only the nearest-neighbor tunneling is considered, which
is described by the parameter 
$J_{a,b}=\int w_{a,b}^{\star}({\bf r}-{\bf r}_{i}) [-\frac{\hbar^2}{2m}\nabla^{2}+V_{l} 
({\bf r})] w_{a,b}({\bf r}-{\bf r}_{j}) \,{\rm d^3} r \;$, 
where $j$ and $i$ are the indices of neighboring sites, 
and $V_{l}({\bf r})=\sum_{\xi=x,y}V_{\xi}^{0}\cos^{2}(k_{\xi}\xi)$ 
is the optical lattice potential (with wavevector ${\bf k}$) 
which we consider to be the same for both species. 
The optical potential for ${\rm K}$ and ${\rm Rb}$ is almost the same for the detunings
corresponding to the wavelength 
$1064{\rm nm}$ of a Nd:Yag laser \cite{Birkl}. 
The same species interact via 
a short range pseudo-potential, which is  described  by the coefficients 
$U_{0a}, U_{0b}$: 
$U_{0a,b}= 4\pi \hbar^{2}a_{\rm K,Rb}
\int |w_{a, b}({\bf r}-{\bf r}_{i})|^{4}/m_{\rm K,Rb} \,{\rm d^3} r$,
where $a_{\rm K}$, $m_{\rm K}$  ($a_{\rm Rb}$, $m_{\rm Rb}$) are the scattering 
length and atomic mass 
of potassium (rubidium). 
The interactions between different species 
are characterized by the coefficient
$U_{ab}=2\pi\hbar^2a_{\rm KRb}
\int |w_{a}({\bf r}-{\bf r}_{i})|^{2} |w_{b}({\bf r}-{\bf r}_{i})|^{2}/m_{ab} 
\,{\rm d^3} r$, where $a_{\rm KRb}$ is the $\rm K-Rb$ scattering length, and 
$m_{ab}$ is the reduced mass for ${\rm K}$ and ${\rm Rb}$ atoms.
The scattering length $a_{\rm KRb}$ has been estimated recently
to be of the order of  $8.6 {\rm nm}$ \cite{Inguscio}, 
that is the repulsive interactions 
between ${\rm K}$ and ${\rm Rb}$ are stronger than those  
between ${\rm K}$-${\rm K}$ and  ${\rm Rb}$-${\rm Rb}$, which are characterized by 
$a_{\rm K}=3.17 {\rm nm}$  and $a_{\rm Rb}=5.24 {\rm nm}$, respectively. 
As a result, it is energetically favorable
for ${\rm K}$ and ${\rm Rb}$ particles to remain immiscible, i.e. to stay at different
lattice sites, rather than in the same ones. 
This situation can be however modified by means of Feshbach resonances, which can 
strongly modify the value of $a_{\rm KRb}$.
We have thus  assumed that the value of $a_{\rm KRb}$
can be reduced to $0.5 \sqrt{a_{\rm K} a_{\rm Rb}}$. 
In that case the two components
acquire a miscible phase.

In the 2D calculations presented below, we have considered 
square lattices up to $80\times 80$ sites, with periodic boundary conditions. 
In the first stage, we demonstrate the 
possibility of performing a transition from a state with two miscible  
superfluids to a Mott insulator phase with exactly one ${\rm Rb}$ and one ${\rm K}$
atom per site. 
To this aim we first obtain the ground state of the system 
using a variational Gutzwiller ansatz \cite{Jaksch98}. 

We approximate the ground-state
wave function by $|\Psi_{MF}\rangle=\prod_{i}|\phi_{i}\rangle$, 
where the product is over all lattice sites. The
functions $|\phi_{i}\rangle$ for each site are expressed in the 
Fock basis, 
$|\phi_{i}\rangle=\sum_{n,m=0}^{\infty}f_{n,m}^{(i)}|n,m\rangle_{i}$, 
where $n$, $m$ indicate the occupation number for 
the ${\rm K}$ and ${\rm Rb}$ atoms, respectively. For numerical reasons
we have assumed that maximally $3$ atoms of each species can be present at 
any lattice site, and checked the validity of this assumption self-consistently.
The coefficients $\{f_{n,m}^{(i)}\}$ are found by minimization
of 
$\langle \Psi_{MF}|H-\sum_{i}[\mu_an_{ai}+\mu_bn_{bi}]|\Psi_{MF}\rangle$, 
where the chemical potentials $\mu_a, \mu_b$
were chosen so that the mean numbers of ${\rm K}$, ${\rm Rb}$ atoms were
equal to the number of lattice sites.

\begin{figure}[hbp]
\begin{center}
{
	\includegraphics[angle=-0,scale=0.25, clip=true]{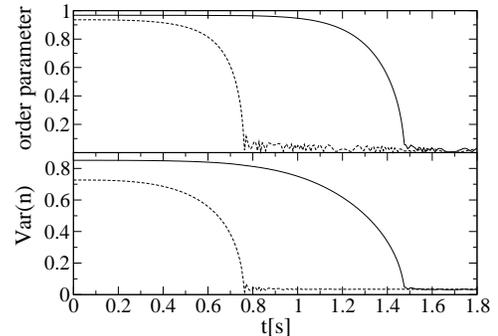}} 

\caption{Dynamical transition from the SF phase of $^{41}{\rm K}$ (solid line)
and $^{87}{\rm Rb}$ (dashed line) atoms to the
MI phase. 
The upper plot shows the value of the 
order parameters $|\langle a_i\rangle|$, $|\langle
b_i\rangle|$ (constant for all lattice sites)
for both species, while the lower 
one depicts the variance ${\rm Var}(n)=\sqrt{\langle n^2\rangle - \langle n
\rangle^2}$ of the on-site occupation.
The lattice potential  
was changed as $V_0(t)= V_{SF} + (V_{MI}-V_{SF}) (t/t_0)^3
\exp(1-(t/t_0)^3)$, where $V_{SF}=4, V_{MI}=19$ in units of recoil energy
of $^{41}{\rm K}$ atoms, $t_0=1.8 s$
for the case of lattice wavelength $1064 {\rm nm}$. 
}
\label{fig1}
\end{center}
\end{figure}

The ground state for weak lattice potentials corresponds to the coexistence of 
superfluids for both atomic species, and it was used as the initial condition  
for subsequent time-dependent calculations. 
This state is characterized by a non-vanishing value of the 
superfluid order parameters $|\langle a_i\rangle|$, $|\langle b_i\rangle|$ 
(Fig. \ref{fig1}). 
To drive the system from superfluid to MI phase we slowly increase
the lattice potential.
This process can be described using a dynamical Gutzwiller method, 
as discussed in Ref. \cite{Jaksch02}. To this aim one allows the 
coefficients $\{f^{(i)}_{n,m}\}$ to be time dependent and to follow the 
dynamics derived from the time dependent variational principle.  
The resulting  equation becomes:
\begin{eqnarray} \label{T}
i\dot{f}^{(i)}_{n,m}&=& \left[\frac{U_{0a}}{2}n(n-1)+\frac{U_{0b}}{2}m(m-1)\right]
f^{(i)}_{n,m}
\nonumber \\
&+&J_{a}\left[\Phi^{a \ast}_{i} \sqrt{n+1} f^{(i)}_{n+1,m}+\Phi^{a}_{i} 
\sqrt{n} f^{(i)}_{n-1,m}\right] \nonumber \\
&+& J_{b} \left[\Phi^{b \ast}_{i} \sqrt{m+1} f^{(i)}_{n,m+1}+\Phi^{b}_{i} 
\sqrt{m} f^{(i)}_{n,m-1}\right] \nonumber \\
&+&U_{ab} n m f^{(i)}_{n,m},
\end{eqnarray}
where 
$$\Phi^{a}_{i}=\sum_{<i,j>} \langle\Psi_{MF}|a_j|\Psi_{MF}\rangle\ ,
\Phi^{b}_{i}=\linebreak \sum_{<i,j>} \langle\Psi_{MF}|b_j|\Psi_{MF}\rangle.$$

When the lattice potential increases, the system follows the changes 
quasi-adiabatically, and the two species enter 
sequentially the Mott phase with 
one atom per site. As shown in Fig. \ref{fig1} this occurs first for the 
heavier ${\rm Rb}$ atoms at $t\approx 0.75 {\rm s}$, 
and then at $t\approx 1.5 {\rm s}$ for ${\rm K}$ atoms.  
The ratio $-U_{0a}/J_a$ ($-U_{0b}/J_b$) 
at the transition point equals approximately $23.5$ ($26$), 
which interestingly 
compares very well with the expected value, $23.2$, 
for a 2D single-component gas. 
Therefore the transition point is rather  unaffected 
by the interactions between different species.

At the end of the process each site contains one ${\rm Rb}$ and one ${\rm K}$ atom, 
with a relative atom number fluctuations less 
than $3\%$. This, however, requires a sufficiently slow modification of the
lattice  potential, typically within a time scale of 2 seconds for
a lattice wavelength $\lambda=1064{\rm nm}$. Note that the time scale 
is provided by the inverse of the recoil energy of the lattice, 
and therefore it 
scales as $\lambda^2$. We note also at this point that
 a transition 3 times faster 
leads to about two times larger number fluctuation in the final state.
We want to stress that during the whole time evolution the
mean number of particles was constant, and equal to the number of lattice
sites, which gives us the most complete analogy, within the Gutzwiller
approach, to the experimental situation.

Having obtained a Mott phase with one Rb and 
one K atom per site, a heteronuclear molecule can be generated at 
each lattice site.  One possible method would be two-color Raman 
photoassociation to make ground state $^{1}\Sigma^+$ or 
$^{3}\Sigma^+$ dimers.  Since heteronuclear species do not have {\it 
gerade/ungerade} symmetry, levels of either symmetry could be formed 
via intermediate excited states of $\Omega=0^+$ or 1 symmetry, where 
$\Omega$ labels projection of electronic plus spin angular momentum 
on the molecular axis.  Since the $\Omega$ states excited at long 
range typically have mixed singlet and triplet character, 
photoassociation to ground $^1\Sigma^+$ levels is even possible for 
collision of two doubly spin-polarized atoms, which collide via the 
ground $^3\Sigma^+$ potential. The $1/R^6$ dependence of the excited 
molecular potential for the case of heteronuclear molecules, in 
contrast to the $1/R^3$ dependence for the homonuclear case, leads to 
excitation at relatively short internuclear separation $R$.    This 
difference may lead to reduced Franck-Condon factors for the first 
exciation step but enhanced factors for the second de-excitation 
step \cite{notka1}.  
In any case, these can be worked out once the potential energy 
curves of the KRb dimer are better known.  A second possible method 
of making ground state molecules is available for heteronuclear 
dimers, namely, use of a pulse of microwave radiation to directly 
associate two colliding atoms in a one-color transition from a 
scattering state to a bound dimer state.  However, the dipole matrix 
elements within the $^{3}\Sigma^+$ state are small,  so this will need fairly high
microwave power.  
Either method will result in an excited vibrational level not far 
below the dissociation limit.  It should be possible to use a 
succession of Raman pulses to transfer the population from such a 
level to the ground $v=0$ vibrational level of the $^1\Sigma^+$ or 
$^3\Sigma^+$ state, in a manner similar to that described by 
Ref.~\cite{Jaksch02}; microwave/infrared pulses may also be feasible 
for this heteronuclear species.  The $^{3}\Sigma^+$ $v=0$ level has a 
small dipole moment of $0.0046$ atomic units (1 atomic unit is 
$8.478\times 10^{-30}$ Cm), whereas it is much larger, 
$0.30$ atomic units, for the $^{1}\Sigma^+$ $v=0$ level 
\cite{julie}.  In the latter case, special care should be taken to 
prevent the heating due to black-body radiation, which could be 
avoided by an appropriate shielding.

After the creation of the K-Rb dimers, the system becomes 
single-component. However,
since  the molecules  have a large 
dipole moment and are considered to be  oriented by an external 
electric field ($\approx$ $100{\rm V/cm}$), we have to include the dipole-dipole interactions 
in our calculations. The corresponding Hamiltonian reads \cite{Goral02}:
\begin{eqnarray} \label{HD}
H&=&J\sum_{<i,j>}b_{i}^{\dagger}b_{j}
+\frac{1}{2}U_{0}\sum_{i}n_{i}(n_{i}-1) \nonumber \\
&+&\frac{1}{2}U_{\sigma_1}\sum_{<i,j>}n_{i}n_{j}
+\frac{1}{2}U_{\sigma_2}\sum_{<<i,j>>}n_{i}n_{j}
+\ldots,
\end{eqnarray}
\noindent where $b_{i}$ is the annihilation operator of a heteronuclear
molecule at the lattice site $i$. The number operator $n_i$, the Wannier basis
$w({\bf r}-{\bf r}_{i})$ and the tunneling
coefficient $J$ for molecules are defined similarly as in Eq. (\ref{H}). 
However, the interaction part significantly differs from that of  
Eq. (\ref{H}). 
The coefficients $U_{\sigma}$ are defined as follows: 
$U_{\sigma}=\int |w({\bf r}-{\bf r}_{i})|^{2} 
V_{int}({\bf r}-{\bf r'})|w({\bf r'}-{\bf r}_{j})|^{2} \,{\rm d^3} r \; 
{\rm d^3} r'$, 
where $\sigma= |{\bf r}_i-{\bf r}_j| |{\bf k}|/4\pi$
is the dimensionless intersite distance.  
In particular, $U_0$  determines the on-site interactions, 
$U_{\sigma_1}$ the nearest-neighbor interactions, $U_{\sigma_2}$ 
the interactions between 
the next-nearest neighbors, etc. Consequently, the respective summations
in  Eq.\ (\ref{HD}) must be carried out over appropriate pairs of sites
which  are marked by $<\dots >$ for the nearest neighbors, 
$<<\dots >>$ for the next-nearest neighbors, etc. 
In the 2D calculations presented below, we have taken into 
account interactions with up to $4$ neighbors 
($\sigma_1=1$, $\sigma_2=\sqrt{2}$, $\sigma_3=2$, $\sigma_4=\sqrt{5}$), 
since the effects of interactions 
of a longer range are negligible. In the case of polarized dipoles 
the interaction potential is
\begin{equation} \label{inter}
V_{int}=d^{2}\frac{1-3\cos^2\theta}{|{\bf r}-{\bf 
r'}|^3}+\frac{4\pi \hbar^{2}a}{M}\delta({\bf r}-{\bf r'}) \; ,
\end{equation}
\noindent where the first part of Eq.\ (\ref{inter}) provides 
the dipole-dipole interaction  
characterized by the dipole moment $d$ and the angle
$\theta$ between the dipole direction and the vector 
${\bf r}-{\bf r'}$. The second part of Eq.\ (\ref{inter}) represents the 
short-range interactions given by the $s$-wave scattering length $a$  
and the molecular  mass $M$. 
We use the dipole moment as $0.3$ a.u. \cite{julie}, and 
assume the $s$-wave scattering length of K-Rb molecule as 
$a=\sqrt{a_{\rm K}a_{\rm Rb}}$\cite{notka}. 


We have checked that the MI phase of these 
heteronuclear molecules is indeed the ground state of the system
for a mean lattice filling factor  equal to 1.
We employ this state as the initial condition for our dynamical calculations of the 
melting of the Mott phase towards a molecular superfluid. 
It is, however, interesting to note, that by itself the MI phase of dipolar particles 
offers promising perspectives as a quantum computation device \cite{qcomp}.
The melting dynamics is 
analyzed by employing again a dynamical Gutzwiller ansatz, but with the 
Hamiltonian (\ref{HD}). During the dynamics we consider a first stage in which 
we reduce the lattice trapping potential up to $4$ in recoil energy units of
$\rm{K-Rb}$ molecules. This first stage allows us to reduce 
the value of $-U_0/J$, but the system remains still in the MI state
mostly due to the large contribution of the repulsive dipolar on-site interactions to
the $U_0$ coefficient. This process can be performed in a relatively fast time scale 
compared to the total duration of the melting process, since the MI 
gap guarantees the adiabaticity at this stage. 

\begin{figure}[hbp]
\begin{center}
{
	\includegraphics[angle=-0,scale=0.25, clip=true]{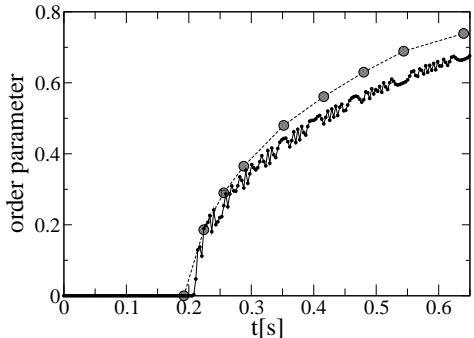}} 

\caption{Melting of $^{41}{\rm K}-^{87}{\rm Rb}$ dimers initially
in the MI  phase. The plot
shows the time evolution of the molecular superfluid order 
parameter $|\langle b_i\rangle|$
(solid line), 
which is the same for all lattice sites.
The dashed line refers to static calculations of the ground state 
of the dipolar molecules placed in the lattice. Note a very sharp transition from
the MI phase to the SF  one, that occurs at $t\approx 0.2 s$.
The frequency of the transverse confinement 
was changed as $\omega_z(t)= \omega_{MI} + 
(\omega_{SF}-\omega_{MI})t/t_0$, where $\omega_{MI}/2\pi=8{\rm kHz}$, 
$\omega_{SF}/2\pi=6{\rm kHz}$, and 
$t_0=0.65 {\rm s}$. 
}
\label{melt}
\end{center}
\end{figure}
Later on, we have observed that the dynamics can be better 
controlled by modifying the transverse confinement. As shown in 
Ref.\ \cite{Goral02,Santos}, the properties of the dipolar gas strongly depend on the 
aspect ratio of the on-site wavefunction. In particular, by reducing the transverse 
confinement, the on-site wavefunction becomes more elongated in transverse
direction. As a consequence,  
$U_0$ decreases \cite{Goral02}, and the system enters into the superfluid phase. 
Fig.\ \ref{melt} shows this process. Due to the non-completely adiabatic evolution 
the gas does not exactly follow the expected stationary result, although very clearly 
a superfluid molecular phase is accomplished. The ratio $-U_0/J \approx 23.2$ 
at which the transition from MI to superfluid equals 
surprisingly  to the value expected for a 2D single-component gas without
a permanent dipole moment. As shown in Fig. \ref{melt} this 
dynamical transition into the superfluid phase occurs for realistic parameters 
in a time scale of $1$s. This transition can be realized, 
$50$ times 
faster, although  
the dynamics of the superfluid order parameter is in such case 
not as smooth as in 
Fig.~\ref{melt}.

In this Letter we have analyzed the formation of a superfluid of polar molecules. 
We  have considered the loading of a BEC of $^{87}$Rb and $^{41}$K atomic species 
into an optical lattice. The modification of the lattice potential produces
the dynamical transition into a MI phase with only one 
atom of each species per site. 
The atoms can then be 
photo-associated on-site. 
Once the polar molecules are created, a molecular MI 
is formed. By reducing the 
lattice potential and modifying the transverse confinement,  
such MI evolves into a molecular superfluid within a feasible time scale.
Eventually, a molecular BEC is formed.  
The mechanism studied in this Letter, provides a perspective 
towards a dipolar BEC.

We acknowledge support from the Alexander von Humboldt Stiftung, the Deutsche 
Forschungsgemeinschaft, EU RTN Network "Cold Quantum Gases", the Polish KBN
(grant no 2 P03B 124 22), ESF PESC Program BEC2000+ and the US Office of Naval
Research. We thank G. Birkl, G. V. Shlyapnikov and  E. Tiesinga for discussions.


\begin{references}
\bibitem{notka1}  Also, the on-site densities ($\approx 10^{15}$ cm$^{-3}$) are $10000$ times 
larger than 
in traditional photoassociation experiments.
\bibitem{notka} The value of $a$ is not known, but does not affect our results, since 
it can always be compensated by transverse confinement, i.e. dipolar forces.
\end{references}
\end{document}